\documentclass[a4paper,12pt]{article}

\usepackage{epsfig}
\usepackage{a4}
\usepackage{latexsym, amssymb} 
\usepackage{amsmath}
\newcommand{\lapprox}{%
\mathrel{%
\setbox0=\hbox{$<$}
\raise0.6ex\copy0\kern-\wd0
\lower0.65ex\hbox{$\sim$}
}}
\newcommand{\gapprox}{%
\mathrel{%
\setbox0=\hbox{$>$}
\raise0.6ex\copy0\kern-\wd0
\lower0.65ex\hbox{$\sim$}
}}

\newcommand{\ba}{\begin{array}}
\newcommand{\ea}{\end{array}}
\newcommand{\bd}{\begin{displaymath}}
\newcommand{\ed}{\end{displaymath}}
\newcommand{\be}{\begin{equation}}
\newcommand{\ee}{\end{equation}}
\newcommand{\bea}{\begin{eqnarray}}
\newcommand{\eea}{\end{eqnarray}}

\def\fb{\, {\rm fb}}

\def\zb{\overline{z}}

\def\tev{\, \, {\rm TeV}}
\def\gev{\, \, {\rm GeV}}

\newcommand{\sig}{\sigma}

\def\thefootnote{\fnsymbol{footnote}}

\begin{document}

\begin{titlepage}

\begin{flushright}
{\small 
KEK-TH-1597\\
IPMU 13-0025}
\end{flushright}

\vspace*{0.2cm}
\begin{center}
{\large {\bf Azimuthal correlation among jets produced in association with a bottom or top quark pair at the LHC}}\\[2cm]  
Kaoru Hagiwara$^{a}$\footnote{kaoru.hagiwara@kek.jp},
Satyanarayan Mukhopadhyay$^{b}$\footnote{satya.mukho@ipmu.jp}\\[0.5cm]  

{\it $^{a}$KEK Theory Center and Sokendai,\\
Tsukuba 305-0801, Japan.\\[0.5cm]
$^{b}$Kavli IPMU (WPI), University of Tokyo,\\
Kashiwa 277-8583, Japan.}\\[2cm]
\end{center}

\begin{abstract}

Angular correlation of jets produced in association with a massive scalar, vector or tensor boson is crucial in the determination of their spin and CP properties. We study jet angular correlations in events with a high mass bottom quark pair or a top quark pair and two jets at the LHC, whose cross-section is dominated by the virtual gluon fusion sub-processes when appropriate kinematic selection cuts (vector-boson fusion cuts) are applied. We evaluate helicity amplitudes for sub-processes initiated by $qq$, $qg$ and $gg$ collisions in the limit where the intermediate gluons are collinear to the initial partons. We first obtain a general expression for the azimuthal angle correlations among the dijets and $t \bar{t}$ or $b \bar{b}$, in terms of the $gg \rightarrow t \bar{t}$ or $b \bar{b}$ helicity amplitudes in the real gluon approximation of the full matrix elements, and find simple analytic expressions in the two kinematic limits, the production of a heavy quark pair near the threshold, and in the relativistic limit where the invariant mass of the heavy quark pair is much larger than the quark mass. For $b\bar{b}+2$~jets we find strong azimuthal angle correlations which are distinct from those expected for events with a CP-even or odd scalar boson which may decay into a $b \bar{b}$ pair. For $t\bar{t}+2$~jets we find that the angular correlations are similar to that of a CP-odd scalar$+2$~jets near the threshold $M_{t \bar{t}}\sim 2 m_t$, while in the relativistic limit they resemble the distribution for $b\bar{b}+2$~jets. These correlations in the standard QCD processes will help establish the experimental technique to measure the spin and CP properties of new particles produced via gluon fusion at the LHC. 
\end{abstract}

\pagestyle{plain}

\end{titlepage}


\setcounter{page}{1}

\renewcommand{\thefootnote}{\arabic{footnote}}
\setcounter{footnote}{0}

\section{Introduction}
\label{sec:intro}
Angular correlations among jets produced in association with scalar bosons and massive gravitons have been the subject of several studies~\cite{HiggsCP,HLM}. It has been found that experimental determination of such correlations can give us important information about the spin and CP-properties of these particles. For example, by studying the distribution of azimuthal angle difference between two tagging jets ($\Delta \phi_{jj}$) in the gluon fusion (GF) production of a scalar ($H$) or pseudo-scalar ($A$) Higgs with two jets, it has been shown that the CP-odd and CP-even Higgs can clearly be discriminated at the LHC. It has also been shown that for jets produced with massive gravitons (G), the $\Delta \phi_{jj}$ distribution is flat while the sum of the azimuthal angles show a characteristic distribution~\cite{HLM}. It was further demonstrated analytically in Ref.~\cite{HLM} that these angular correlations arise from the quantum interference of different helicity states of the intermediate vector bosons, which can either be a weak boson or a gluon. Another critical observation made in Ref.~\cite{HLM} is that the amplitudes for $H/A/G+2$~jets production can be approximated by those of the vector boson fusion (VBF) sub-processes when a specific set of kinematic cuts are applied, where the two tagging jets are demanded to be in the opposite hemispheres of the detector and also to be well separated in rapidity (the so called VBF cuts), and in addition their transverse momenta are subjected to a slicing cut, restricting them to be sufficiently low as compared to the H/A/G masses. We note in passing that another complementary method of determining the spin and CP information of new particles is from their decay patterns (for example, the decay of a Higgs boson to a pair of gauge bosons), which has also been widely discussed in the literature~\cite{HiggsCP,HLM,Nelson}.

In this paper, we study heavy quark pair $Q \overline{Q}$ ($Q=b,t$) production at the LHC in association with two jets, and study the azimuthal angle correlations among these jets. Our goal is to predict the jet angular correlations for the SM processes by using the same approximation, so that the experimental technique to measure the new particle properties via initial state radiation patterns can be established using the ample SM processes. We first calculate the helicity amplitudes for gluon fusion processes initiated by $qq,qg,gg$ initial states at the LHC leading to $Q \overline{Q}+2$~jets production. Here, we work in the limit in which the intermediate gluons are collinear to the initial partons. We work out the angular correlations using the formalism of Ref.~\cite{HLM}, and then verify our approximate analytical results in several kinematic regions by comparing them with the exact parton level matrix elements in the tree level.

This paper is organized as follows. In section 2 we briefly review the helicity formalism used to obtain our analytical results. In section 3 we give the spin summed matrix-element squared for $Q\overline{Q}+2$~jets production via gluon fusion in the collinear limit for the exchanged gluons, and then analyze the results in two different kinematic limits, namely that of the $t \bar{t}$ production threshold ($Q=t$) in which the invariant mass of the $t\overline{t}$ pair, $M_{t\overline{t}}$, is very close to $2m_t$, and that of the relativistic limit where $M_{Q\overline{Q}}$ is much higher than $2m_Q$, for $Q=b$ and $t$. In section 4 we show the exact numerical results for these angular correlations at parton level, where the expected azimuthal angle correlations are reproduced once appropriate final state selection cuts are applied. Finally, we summarize our findings in section 5.

\section{Helicity formalism : brief review}
\begin{figure}[h]
\begin{center}
\centerline{\epsfig{file=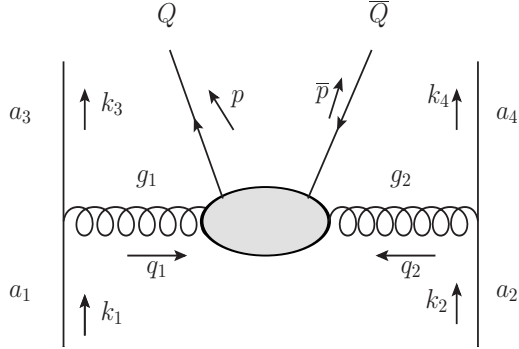,width=7.5cm,height=5cm,angle=-0}}

\caption{Schematic Feynman diagram for the parton level $Q \overline{Q}jj$ production process in the on-shell gluon approximation, where $a_i$ denote quarks, anti-quarks or gluons. The four-momenta of each particle are also shown along the particle lines.}
\label{fig:Fig-feynman}
\end{center}
\end{figure}

The helicity amplitude formalism for vector boson fusion processes has been discussed in detail in Ref.~\cite{HLM} in the context of production and decay of a heavy particle in association with two jets. We follow the same notations and conventions in this paper. The inclusive process,
\begin{equation}
pp\rightarrow Q \overline{Q} jj + \mbox{anything}~,
\end{equation}
can proceed, via gluon fusion, through the sub-processes $a_1 a_2 \rightarrow a_3 a_4 g^* g^*\rightarrow a_3 a_4 Q \overline{Q}$, where $a_1a_2$ is one of the three possible initial states $\{qq,qg,gg\}$, $g^*$ is a $t$-channel intermediate gluon and $q$ stands for a quark or anti-quark of any flavour: 
\begin{subequations}
\label{eq1}
\begin{align}
qq &\rightarrow qqQ\overline{Q} \label{eq11}\\
qg &\rightarrow qgQ\overline{Q} \label{eq12}\\
gg &\rightarrow ggQ\overline{Q}~. \label{eq13}
\end{align}
\end{subequations}

In addition to gluon fusion, the above sub-processes receive contributions from all other diagrams at the same order in perturbation theory to make a gauge-invariant physical amplitude. However, in this and the next section, we shall consider only the gluon fusion diagrams. As mentioned earlier, it has been demonstrated in Ref.~\cite{HLM} that after applying the VBF selection cuts, the GF contribution can be made to dominate the total cross-section. We shall also show the validity of this approach by first determining analytically the distributions predicted by the GF diagrams in various kinematic limits, and then comparing them with the exact matrix elements.

To begin with, we define a common set of kinematic variables for the $Q \overline{Q}jj$ sub-processes in Eqn.~\ref{eq1} as

\begin{align}
 a_1(k_1,\sig_1)+a_2(k_2,\sig_2) &\rightarrow a_3(k_3,\sig_3)+a_4(k_4,\sig_4)+Q(p,\sigma) + \overline{Q}(\bar{p},\bar{\sigma}),
\label{qq_qqx}
\end{align}
The helicities of the quarks or anti-quarks have the value $\sigma_i/2$ while for on-shell gluons they take the values $\sigma_i=\pm 1$. The helicity amplitudes for this process can be written as 
\begin{align}
\mathcal{M}_{\sigma_1 \sigma_3,\sigma_2\sigma_4}^{\sigma \bar{\sigma}} &= J^{\mu_1^\prime}_{a_1 a_3} (k_1,k_3;\sigma_1,\sigma_3) J^{\mu_2^\prime}_{a_2 a_4} (k_2,k_4;\sigma_2,\sigma_4) \nonumber \\
&   \times D^{g_1}_{\mu_1^\prime \mu_1} (q_1) D^{g_2}_{\mu_2^\prime \mu_2} (q_2) \hat{\mathcal{M}}_{Q\overline{Q} g_1 g_2}^{\mu_1 \mu_2} (q_1,q_2,p,\bar{p};\sigma,\bar{\sigma}) 
\label{Eq:ampl}
\end{align}
where the external quark or gluon currents are denoted by $J^{\mu_i^\prime}_{a_i a_{i+2}} $, and the gluon propagator is given by
\begin{equation}
D^{g_i}_{\mu_i^\prime \mu_i} (q_i) = \frac{-g_{\mu_i^\prime \mu_i} }{q_i^2} = \frac{1}{q_i^2}\sum_{\lambda_i=\pm 1} \epsilon_{\mu_i^\prime}^* (q_i,\lambda_i)\epsilon_{\mu_i} (q_i,\lambda_i)~,
\end{equation}
for the conserved currents
\begin{equation}
q_{\mu_1^\prime}J^{\mu_1^\prime}_{a_1 a_3} (k_1,k_3;\sigma_1,\sigma_3)=q_{\mu_2^\prime}J^{\mu_2^\prime}_{a_2 a_4} (k_2,k_4;\sigma_2,\sigma_4)=0. \label{Eq:curr-consv}
\end{equation}
We note in passing that the conditions in Eqn.~\ref{Eq:curr-consv} are satisfied not only for massless quark currents but also for the gluonic currents, in the processes (2b) and (2c) in the light-cone gauge~\cite{HLM}. The real gluon approximation to the amplitudes (Eqn.~\ref{Eq:ampl}) is obtained by replacing the off-shell $g^*g^* \rightarrow Q \overline{Q}$ amplitudes by the on-shell, and hence gauge-invariant $gg \rightarrow Q \overline{Q}$ amplitudes
\begin{equation}
\epsilon_{\mu_1}(q_1,\lambda_1)\epsilon_{\mu_2}(q_2,\lambda_2) \hat{\mathcal{M}}_{Q\overline{Q} g_1 g_2}^{\mu_1 \mu_2} (q_1,q_2,p,\bar{p};\sigma,\bar{\sigma}) \xrightarrow{q_i^2\rightarrow 0} (\mathcal{M}^{Q \overline{Q}}_{g_1 g_2})_{\lambda_1 \lambda_2}^{\sigma \bar{\sigma}}~,
\end{equation}  
while keeping the four-momenta of the $Q \overline{Q}$ system and the orientation of the colliding virtual gluon momenta in the $Q\overline{Q}$ rest-frame. The full amplitudes for the process (3) are now approximated as 
\begin{equation}
\mathcal{M}_{\sigma_1 \sigma_3,\sigma_2\sigma_4}^{\sigma \bar{\sigma}} \approx \frac{1}{q_1^2 q_2^2} \sum_{\lambda_i=\pm 1} (\mathcal{J}^{g_1}_{a_1 a_3})_{\sigma_1 \sigma_3}^{\lambda_1}(\mathcal{J}^{g_2}_{a_2 a_4})_{\sigma_2 \sigma_4}^{\lambda_2}(\mathcal{M}^{Q \overline{Q}}_{g_1 g_2})_{\lambda_1 \lambda_2}^{\sigma \bar{\sigma}}~,
\label{eq-gen-form}
\end{equation}
where the incoming current amplitudes are given by

\begin{equation}
(\mathcal{J}^{g_i}_{a_i a_{i+2}})_{\sigma_i \sigma_{i+2}}^{\lambda_i} =  J^{\mu}_{a_i a_{i+2}} (k_i,k_{i+2};\sigma_i,\sigma_{i+2}) \epsilon_{\mu} (q_i,\lambda_i)^*
\end{equation}
It has been shown in Ref.~\cite{HLM} that the helicity dependent phases of the above current amplitudes reproduce azimuthal angle correlations among the jets while their magnitudes give the well-known DGLAP splitting functions when squared and summed over helicities.

Following Ref.~\cite{HLM}, we define the reduced current amplitudes $\mathcal{\hat{J}}^{\lambda_i}_{\sigma_i \sigma_{i+2}}$ as follows

\begin{equation}
(\mathcal{J}^{g_i}_{a_i a_{i+2}})_{\sigma_i \sigma_{i+2}}^{\lambda_i} = \sqrt{2} g_{g_i}^{a_i a_{i+2}} Q_i \mathcal{\hat{J}}^{\lambda_i}_{\sigma_i \sigma_{i+2}}~,
\end{equation}
where $Q_i=\sqrt{-q_i^2}$ and the coupling factor $g_{g_i}^{a_i a_{i+2}}$ is given by $g_s T^a$ for quark or anti-quark currents, and $g_s f_{abc}$ for gluon currents. Here, $g_s = \sqrt{4 \pi \alpha_s}$ is the QCD coupling constant, while we suppress the colour indices in the helicity amplitudes of Eqn.~\ref{Eq:ampl} for brevity. The reduced current amplitudes $\mathcal{\hat{J}}^{\lambda_i}_{\sigma_i \sigma_{i+2}}$ have been computed and tabulated in Ref.~\cite{HLM}, to which we refer the reader for details. However, it might be worth recalling some of the salient features of these reduced current amplitudes. First of all, the reduced amplitudes for anti-quark currents are same as those of quark currents. And since we neglect light quark masses, by chirality conservation, the helicity-flip amplitudes will be zero for quark currents, i.e., $\mathcal{\hat{J}}^{\lambda_i}_{\sigma -\sigma}=0$, while this is not the case for gluon current amplitudes. The existence of helicity-flip amplitudes in gluon currents give rise to certain special features in the amplitudes for $Q \overline{Q}+2$~jets involving them. 

It is useful to look into the current amplitudes from the point of view of parton branching~\cite{parton-branch},  whereby the outgoing partons are emitted collinearly. In the current amplitudes $f \rightarrow f g^*$ or $g \rightarrow g g*$, if $z$ is the energy fraction of the initial parton that is carried away by the off-shell gluon, then one can write the current amplitudes in terms of the energy fractions of the two t-channel gluons, $z_1$, $z_2$, and the azimuthal angles of the outgoing partons $\phi_1$ and $\phi_2$, in the frame in which the t-channel gluons are taken as propagating along the $ \pm z$-axis.

\subsection{Helicity amplitudes for $gg \rightarrow Q \overline{Q}$}
\label{sec:gg-QQ}
Let us first fix the reference frame in which we shall be working. As mentioned above, we take this frame to be the centre of mass frame of the two t-channel gluons producing the $Q \overline{Q}$ pair, which  is also the rest frame of the $Q \overline{Q}$ pair. In this frame, the two gluons can be taken to be propagating in opposite directions along the $z$-axis with energy $E$ each. Therefore, the scattering process,
\begin{equation}
g(q_1,\lambda_1,a_1)+g(q_2,\lambda_2,a_2) \rightarrow Q(p,\sigma,i)+\overline{Q}(\bar{p},\bar{\sigma},\bar{i})~,
\end{equation}
can be assumed to be taking place in the $x-z$ plane. Here, $a_1, a_2$ denote the colour indices for the gluons and $i,\bar{i}$ denote the colour indices for the quark and the anti-quark. The four momenta of the gluons, $Q$ and $\overline{Q}$ are given by
\begin{align}
q_1^\mu &=E(1,0,0,1) \nonumber \\
q_2^\mu &=E(1,0,0,-1) \nonumber \\
p^\mu &= E(1,\beta \sin \theta,0,\beta\cos\theta) \nonumber \\
\bar{p}^\mu &= E(1,-\beta \sin \theta,0,-\beta\cos\theta)
\end{align}
where $\beta$ denotes the velocity of $Q$ in the $Q \overline{Q}$ rest frame and is given by $\beta = \sqrt{1-\frac{4m^2}{M_{Q \overline{Q}}^2}}$, and $M_{Q \overline{Q}}$ is the invariant mass of the $Q \overline{Q}$ system.

The helicity amplitudes for the process $gg \rightarrow Q \overline{Q}$ can be expressed in the following form: 
\begin{equation}
(\mathcal{M}^{Q \overline{Q}}_{g_1 g_2})_{\lambda_1 \lambda_2 a_1 a_2}^{\sigma \bar{\sigma}i \bar{i}} = g_s^2 \left[\frac{1}{2} \{T^{a_1},T^{a_2}\}_{i,{\overline{i}}} \mathcal{\hat{M}}_{\lambda_1 \lambda_2}^{\sigma \bar{\sigma}} + \frac{1}{2} [T^{a_1},T^{a_2}]_{i,{\overline{i}}} \mathcal{\hat{N}}_{\lambda_1 \lambda_2}^{\sigma \bar{\sigma}}\right]~.
\label{eq-amp-ggQQ}
\end{equation}
One can further simplify Eqn.~\ref{eq-amp-ggQQ} by noting the following relation 
\begin{equation}
\mathcal{\hat{N}}_{\lambda_1 \lambda_2}^{\sigma \bar{\sigma}} = \mathcal{\hat{M}}_{\lambda_1 \lambda_2}^{\sigma \bar{\sigma}} \times \beta \cos \theta~.
\end{equation}
\pagebreak
Finally, the helicity amplitudes $\mathcal{\hat{M}}_{\lambda_1 \lambda_2}^{\sigma \bar{\sigma}}$, for different helicity combinations of the heavy quarks and gluons, are given by

\begin{align}
\mathcal{\hat{M}}_{\lambda -\lambda}^{\sigma -\sigma} &= 2 \beta \sin \theta (\sigma \lambda + \cos \theta)/(1-\beta^2 \cos^2 \theta) \nonumber \\
\mathcal{\hat{M}}_{\lambda \lambda}^{\sigma \sigma} &= -2 \sqrt{1-\beta^2} (\lambda + \sigma \beta)/(1-\beta^2 \cos^2 \theta) \nonumber \\
\mathcal{\hat{M}}_{\lambda -\lambda}^{\sigma \sigma} &= 2 \beta \sqrt{1-\beta^2}\sigma \sin^2 \theta /(1-\beta^2 \cos^2 \theta) \nonumber \\
\mathcal{\hat{M}}_{\lambda \lambda}^{\sigma -\sigma} &= 0 ~.
\label{eq-hel-amp}
\end{align}

\section{Azimuthal angle correlations: analytical results}
\label{analytic}
The squared matrix element (averaged over initial state and summed over final state colour (c) and spin (s)) can be obtained by using Eqns.~\ref{eq-gen-form},~\ref{eq-amp-ggQQ},~\ref{eq-hel-amp}, and the current amplitudes $\mathcal{J}^{g_i}_{a_i a_{i+2}}$ as tabulated in Ref.~\cite{HLM}. We find that for all of the different collision processes initiated by $qq,qg$ and $gg$, the squared amplitude can be cast in the following form: 
\begin{align}
\overline{\sum_{s,c}} |\mathcal{M}_{\sigma_1 \sigma_3,\sigma_2\sigma_4}^{\sigma \bar{\sigma}}|^2 &= \frac{16g_s^8 C}{Q_1^2 Q_2^2} \frac{(7/3+3 \beta^2 \cos^2 \theta)}{(1-\beta^2 \cos^2 \theta)^2} [\lbrace 1+\beta^2 \sin^2 \theta(2-\beta^2 \sin^2 \theta)-\beta^4 \rbrace F_0 \nonumber \\
&  + 4 \beta^2 (1-\beta^2) \sin^2 \theta \{F_1 \cos (2 \phi_1) + F_2 \cos (2 \phi_2)\} \nonumber \\
&  -4 (1-\beta^2)^2 F_3 \cos 2(\phi_1-\phi_2)-4 \beta^4 \sin^4 \theta F_4 \cos 2(\phi_1+\phi_2)]
\label{eq-Msq}
\end{align}
where the colour factor $C$ comes from the quark and gluon current amplitudes, averaged over initial colour, and is given by $C=1/36,1/16$ and $9/64$ for $qq,qg$ and $gg$ sub-processes respectively. The co-efficients $F_i$'s in the above expression are functions of $z_i$, the energy fractions of the initial partons carried away by the t-channel intermediate gluons. For notational convenience, let us also define $\bar{z_i}=1-z_i$. Here $\phi_1$ and $\phi_2$ are the azimuthal angles of the two outgoing partons (jets) in the $Q \overline{Q}$ rest frame whose polar axis is chosen along the virtual gluon momentum as defined in sub-section~\ref{sec:gg-QQ}. In the collinear limit, these azimuthal angles reduce to the standard ones in the Lab frame with polar axis along the beam direction. For definiteness, we take the heavy quark Q (Q=b or t) momentum to have zero azimuthal angle. 

The general expressions for $F_0$, which determines the total cross-section, for the $qq$, $qg$ and $gg$ initiated sub-processes are as follows:
\begin{align}
 F_0[qq] &= \frac{(1+\overline{z_1}^2)}{z_1^2} \frac{(1+\overline{z_2}^2)}{z_2^2} \nonumber \\
 F_0[qg] &= \frac{(1+\overline{z_1}^2)}{z_1^2}\frac{(1+z_2^4+\zb_2^4)}{z_2^2 \zb_2} \nonumber \\
 F_0[gg] &= \frac{(1+z_1^4+\zb_1^4)}{z_1^2 \zb_1} \frac{(1+z_2^4+\zb_2^4)}{z_2^2  \zb_2}~.
 \label{eq-F0}
\end{align}
The function $F_1$  is found to be the same for the $qg$ and $gg$ initiated processes, and is given by,
\begin{align}
F_1[qq] &= \frac{\overline{z_1}}{z_1^2} \frac{(1+\overline{z_2}^2)}{z_2^2}  \nonumber \\
F_1[qg/gg] &= \frac{\zb_1}{z_1^2} \frac{(1+z_2^4+\zb_2^4)}{z_2^2 \zb_2}~.
\label{eq-F1}
\end{align}
Similarly, the function $F_2$ is the same for the $qq$ and $qg$ initial states, and is as follows:
\begin{align}
F_2[qq/qg] &= \frac{\overline{z_2}}{z_2^2} \frac{(1+\overline{z_1}^2)}{z_1^2 } \nonumber \\
F_2[gg] &= \frac{\zb_2}{z_2^2}\frac{(1+z_1^4+\zb_1^4)}{z_1^2 \zb_1}~.
\label{eq-Fqg}
\end{align}
The $z_i$-dependent co-efficients of the terms proportional to $\cos 2 (\phi_1-\phi_2)$ and $\cos 2 (\phi_1+\phi_2)$, namely $F_3$ and $F_4$, which will be seen to determine the azimuthal correlations in the threshold and the relativistic limits respectively, are found to be equal to each other (i.e., $F_3=F_4$), and are also remarkably the same for all three sub-processes,
\begin{equation}
F_{3,4}[qq/qg/gg] = \frac{\zb_1}{z_1^2} \frac{\zb_2}{z_2^2}~.
\label{eq-Fgg}
\end{equation}
We see that there is an enhancement in all the amplitudes in the $z_1 \rightarrow 0$ and $z_2 \rightarrow 0$ limits, i.e., when the t-channel intermediate gluons become soft. Furthermore, in the amplitudes involving gluon currents (i.e., the $qg$ and $gg$ initiated processes), there is an additional singularity as $z_i \rightarrow 1$ (i.e., $\zb_i \rightarrow 0$). This happens because not only the intermediate gluons can become soft, but now an outgoing gluon jet can become soft as well. Since the form of $F_3$ and $F_4$ are identical for all of the $qq,qg$ and $gg$ collision processes, and these functions determine the angular distributions in the two kinematic limits to be considered next, the only difference that will arise is from $F_0$. We shall see that the ratio of $F_{3}$ or $F_4$ to $F_0$ determines how strong the azimuthal correlations will be. 
\subsection{Threshold and relativistic limits}

We now explore the two different kinematic limits in which one obtains characteristic strong azimuthal angle correlations that can be probed at the LHC. The first one is when the $Q \overline{Q}$ pair is produced near the threshold, i.e., the invariant mass $M_{Q \overline{Q}} \approx 2m_Q$, and therefore, $\beta \rightarrow 0$. We can then express the squared matrix element in Eqn.~\ref{eq-Msq} in this simple form:

\begin{equation}
 \overline{\sum_{s,c}} |\mathcal{M}_{\sigma_1 \sigma_3,\sigma_2\sigma_4}^{\sigma \bar{\sigma}}|^2  \xrightarrow{\beta \rightarrow 0} \frac{16g_s^8 C}{Q_1^2 Q_2^2} \frac{7 F_0}{3} \left[1-\frac{4 F_3}{F_0} \cos 2 (\phi_1-\phi_2)\right]
\label{Eq:thres}
\end{equation}
From this equation, we can see that the differential distribution in $(\phi_1-\phi_2)$ will show a strong suppression at $(\phi_1-\phi_2)=0,\pm \pi $, while it will show an enhancement at $(\phi_1-\phi_2)=\pm \pi/2 $ (we use the convention in which the azimuthal angles are in the interval $[-\pi,\pi]$). Note that, in Eqn.~\ref{Eq:thres}, in the soft-limit for the t-channel gluons ($z_i \rightarrow 0$), the combination $4F_3/F_0 \rightarrow 1$. In the next section we shall see that by studying the threshold production of top quark pair in $t \bar{t}+2$~jets events at the LHC, we do observe these features in the corresponding differential distribution for $(\phi_1-\phi_2)$. We would like to point out that in the production of a CP-odd scalar boson in association with 2 jets one expects a similar distribution in $(\phi_1-\phi_2)$  .

The second kinematic region of interest is the relativistic limit in which the $Q \overline{Q}$ pair invariant mass is very high, such that the masses of the heavy quarks can be ignored. In this limit, as $m \ll E$, $\beta \rightarrow 1$. Therefore, we can express the squared matrix element in Eqn.~\ref{eq-Msq} as follows:
\begin{equation}
 \overline{\sum_{s,c}} |\mathcal{M}_{\sigma_1 \sigma_3,\sigma_2\sigma_4}^{\sigma \bar{\sigma}}|^2 \xrightarrow{\beta \rightarrow 1} \frac{16g_s^8 C}{Q_1^2 Q_2^2} \frac{(7/3+3 \cos^2 \theta)(1+\cos^2 \theta)}{\sin^2 \theta} F_0  \left[1-\frac{\sin^2\theta}{1+\cos^2\theta}\frac{4 F_4}{F_0} \cos 2 (\phi_1+\phi_2)\right]
\label{Eq:relat}
 \end{equation}
From the above expression we see that in the relativistic limit, the differential distribution in the sum of azimuthal angles, $(\phi_1+\phi_2)$, will show a strong suppression at $(\phi_1+\phi_2)=0,\pm \pi$, while it will show an enhancement at $(\phi_1+\phi_2)=\pm \pi/2$. In the next section, we shall study $Q \overline{Q}+2$~jets ($Q=b,t$) processes at the LHC by demanding a high invariant mass for the $Q \bar{Q}$ pair, and shall see that we are able to observe these features in the differential distribution for $(\phi_1+\phi_2)$.
\section{Numerical results}

We now discuss the results for exact numerical evaluation of the differential distributions for $(\phi_1-\phi_2)$ and $(\phi_1+\phi_2)$ in the $t \bar{t}+2$~jets and $b \bar{b}+2$~jets processes. Here, in addition to the GF diagrams, all other diagrams at the tree level have also been taken into account. Using suitably devised cuts, we show that we are able to observe the angular correlations in different kinematic regions, as obtained using only the GF contribution in the previous section. We also compare the distributions and total cross-sections obtained by using the approximate matrix elements (presented in section~\ref{analytic}, Eqns.~\ref{eq-Msq}-\ref{eq-Fgg}) with the exact ones. For exploring the threshold limit, we concentrate on the process $t \bar{t}+2$~jets, with $M_{t \bar{t}}$ close to $2m_t$. For the relativistic limit, in which $m \ll E$, we consider both the production of $b \bar{b}+2$~jets and $t \bar{t}+2$~jets with a high invariant mass of the bottom or top pair. 

We present our results for LHC with a centre of mass energy of $8 \tev$. The exact matrix elements were calculated with the help of {\tt MadGraph 5}~\cite{Madgraph} at parton level. Subsequently, both the exact and approximate matrix elements were integrated over the relevant four-body phase-space using the {\tt BASES}~\cite{BASES} Monte-Carlo integration package. Note that, we do not include the effects of parton shower in this study. We have used the {\tt CTEQ6L1}~\cite{cteq} parton distribution functions with LO running of $\alpha_s$, and $\alpha_s (M_Z) = 0.13$. For the factorization scale we choose the minimum value of transverse momenta demanded for the tagging jets (${p_T^j}_{min}=20 \gev$). The strong coupling constant $\alpha_s$ is evaluated at the corresponding minimum transverse energy of the final state partons or heavy quarks, i.e., for the $t \bar{t}+2$~jets process, we take $\alpha_s^4=\alpha_s(m_t)^2 \alpha_s({p_T^j}_{min})^2$, while for $b \bar{b}+2$~ jets it is taken to be $\alpha_s^4=\alpha_s({E_T^b}_{min})^2 \alpha_s({p_T^j}_{min})^2$. In our analysis, ${E_T^b}_{min}$ is taken as $100 \gev$ (see Cut-3 below). The values of top and bottom quark masses used are $173 \gev$ and $4.7 \gev$ respectively~\cite{PDG}. 

To start with, we impose a set of selection cuts on the final-state partons (jets):
\vspace{0.3cm}
\begin{center}
{\bf Cut-1}:~~~$20 \gev \leq p_T^{j} \leq 60 \gev$,  ~~$E_j \geq 250 \gev$,  ~~$|\eta_{j}| \leq 5$, ~~$\Delta R_{j_1j_2}\geq 0.6$
\end{center}
\vspace{0.3cm}
Here $p_T^j$, $E_j$ and $\eta_j$ denote the transverse momenta, energy and pseudorapidity of the jets respectively, while $\Delta R_{j_1j_2} \left(=\sqrt{(\eta_{j1}-\eta_{j_2})^2+(\phi_{j1}-\phi_{j_2})^2}\right)$ defines the separation between the jets in the pseudorapidity-azimuthal angle plane. We have subjected the tagging jets to a slicing $p_T$ cut, and have also demanded a high value for the minimum energy of the jets. Since we want to isolate the contribution of the GF diagrams to the full process, we also need to impose the so-called VBF cuts, which require the two tagging jets to reside in the opposite hemispheres of the detector and to be well separated in pseudorapidity. The VBF cuts imposed in our study are as follows:
\vspace{0.3cm}
\begin{center}
{\bf Cut-2}:~~~$\eta_{j_1}>0>\eta_{j_2}$,  ~~${\Delta \eta}_{jj} = \eta_{j_1} - \eta_{j_2} > 4$.
\end{center}
\vspace{0.3cm}
In addition to the above cuts on the tagging jets, we also impose a set of cuts on the $t$ and $b$ quarks. We demand a minimum transverse energy ($E_T=\sqrt{m^2+p_T^2}$) for the $b$ quarks, such that $E_T^{b,\bar{b}}$  is always higher than the tagging jet $p_T$'s. Also, the $t$ and $b$ quarks are required to be in the central region of the detector. These cuts are summarized below:
\vspace{0.3cm}
\begin{center}
{\bf Cut-3}:~~~$E_T^{b,\bar{b}} \geq 100 \gev$,  ~~$|\eta|_{b,\bar{b},t,\bar{t}} \leq 2.5$ 
\end{center}
\vspace{0.3cm}
\begin{figure}[h]
\begin{center}
\centerline{\epsfig{file=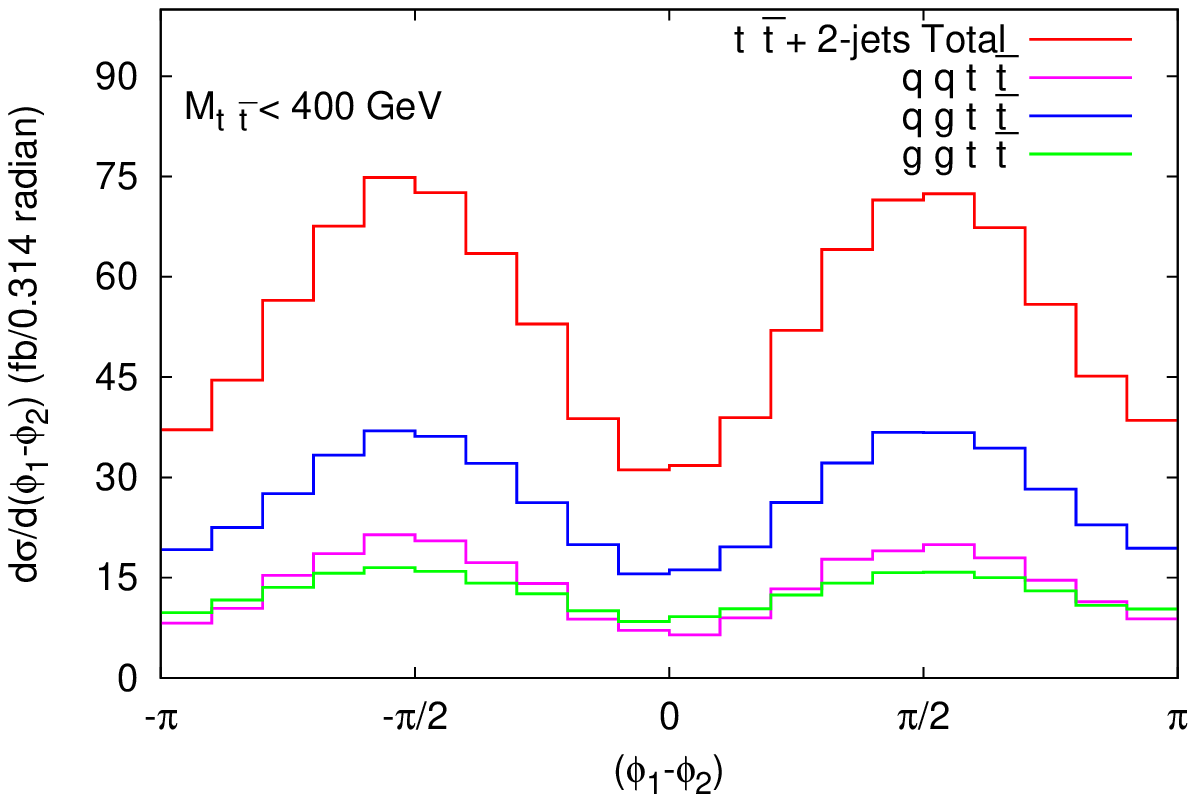,width=8.5cm,height=7cm,angle=-0}
\epsfig{file=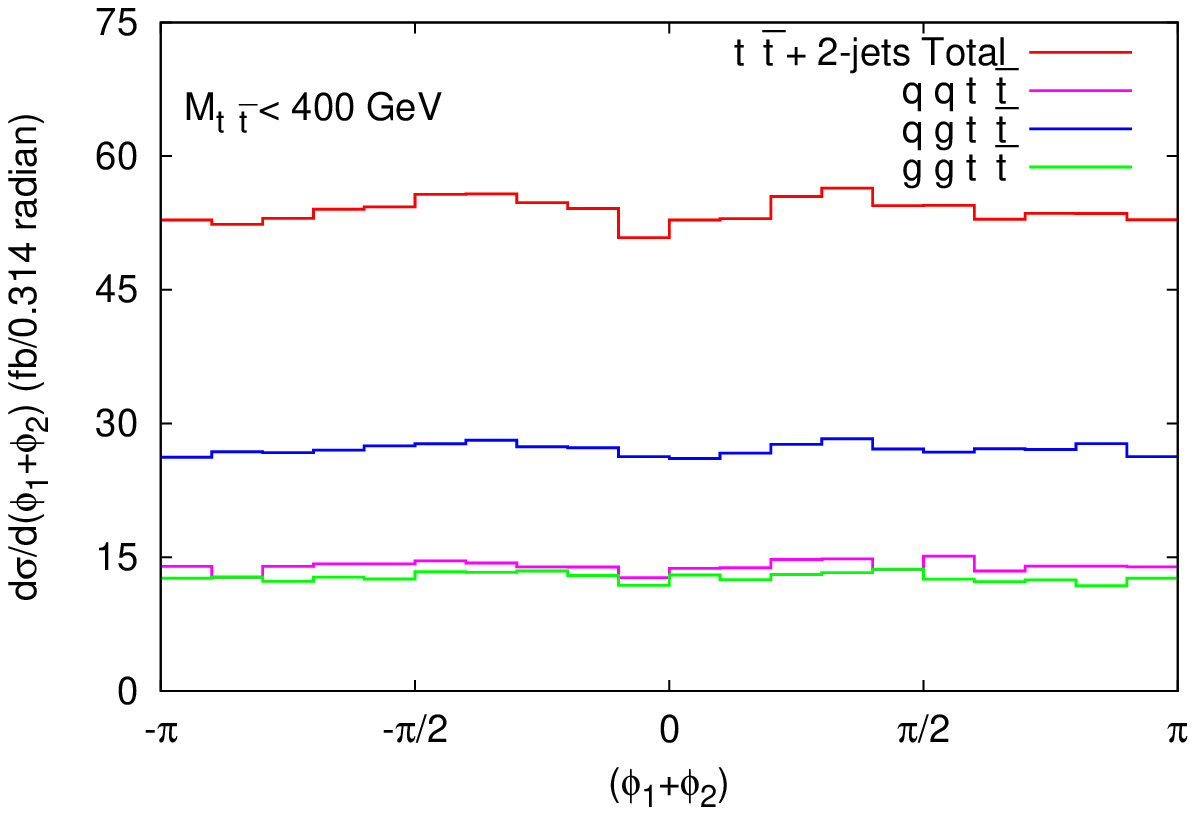,width=8.5cm,height=7cm,angle=-0}}

\caption{Differential distribution of $(\phi_1-\phi_2)$ (left panel) and $(\phi_1+\phi_2)$ (right panel) for $t \bar{t}+2$~jets in the three sub-processes initiated by $qq,qg$ and $gg$ and the sum of all sub-processes at the $\sqrt{s}= 8 \tev$ LHC after Cuts 1, 2 and 3 as described in the text. The invariant mass of the $t \bar{t}$ pair has been demanded to be $M_{t\bar{t}} < 400 \gev$.}
\label{fig:Fig-1}
\end{center}
\end{figure}

In Figure~\ref{fig:Fig-1}, we show the differential distributions obtained using the exact matrix elements for both the azimuthal angle difference of the two tagging jets, $(\phi_1-\phi_2)$, and the sum of the azimuthal angles $(\phi_1+\phi_2)$ in $t \bar{t}+2$~jets events, for the sub-processes initiated by $qq, qg$ and $gg$. In addition, we show the sum of all sub-process contributions, too. In order to obtain the azimuthal angle distributions in the threshold limit, we need to ensure that the invariant mass of the $t \bar{t}$ system, $M_{t\bar{t}}$, should be very close to $2 m_t$. Therefore, in this figure, we have imposed an invariant mass cut of $M_{t\bar{t}} < 400 \gev$. As explained in sub-section 3.1, we indeed see a distribution resembling the behaviour predicted by Eqn.~\ref{Eq:thres}. In particular, we see the expected suppression at $(\phi_1-\phi_2)=0,\pm \pi $, and enhancement at $(\phi_1-\phi_2)=\pm \pi/2 $. On the otherhand, the $\phi_1+\phi_2$ distribution is observed to be rather flat, as expected in this region of $M_{t\bar{t}}$. This agreement between the analytic approximation and the exact numerical evaluation shows the usefulness of the VBF cuts in selecting the GF contribution to the total cross-section. In order to estimate the feasibility of studying the angular distributions in the $t \bar{t}+2$~jets process at the $8 \tev$ LHC, we show the cross-sections after cuts 1, 2 and 3, and various values of the $M_{t\bar{t}}$ cut in Table~\ref{Mtt}. As we can infer from this table, the total cross-section from all sub-processes taken together is considerable, such that with an expected integrated luminosity of $\sim 20 \fb^{-1}$, one can have sufficient number of events to study the azimuthal angle correlations at the 8 TeV run of LHC. 
\begin{table}[htb]
\begin{center}
\begin{tabular}{|l|c c c c|}
\hline
$M_{t \bar{t}}$ cut & $\sigma_{qqt\bar{t}}$ (fb) &$\sigma_{qgt\bar{t}}$ (fb) &$\sigma_{ggt\bar{t}}$ (fb) & $\sigma_{t\bar{t}+2~jets}$ (fb) \\
\hline 
No $M_{t \bar{t}}$ cut     &655.80  &1199.09 &529.43 &2384.32 \\
$M_{t \bar{t}} < 400$ GeV  &87.91  &170.22 &80.12 &338.25  \\
$M_{t \bar{t}} > 600$ GeV  &195.62  &334.58 &137.42 & 667.62 \\
\hline 
\end{tabular}
\end{center}
\caption{\label{Mtt} Cross-sections of various sub-processes contributing to $t \bar{t}+2$~jets at 8 TeV LHC after Cuts 1, 2 and 3 and different values of the $M_{t \bar{t}}$ cut. We also show the total $t \bar{t}+2$~jets cross-section after these cuts.}
\end{table}
\begin{figure}[h!]
\begin{center}
\centerline{\epsfig{file=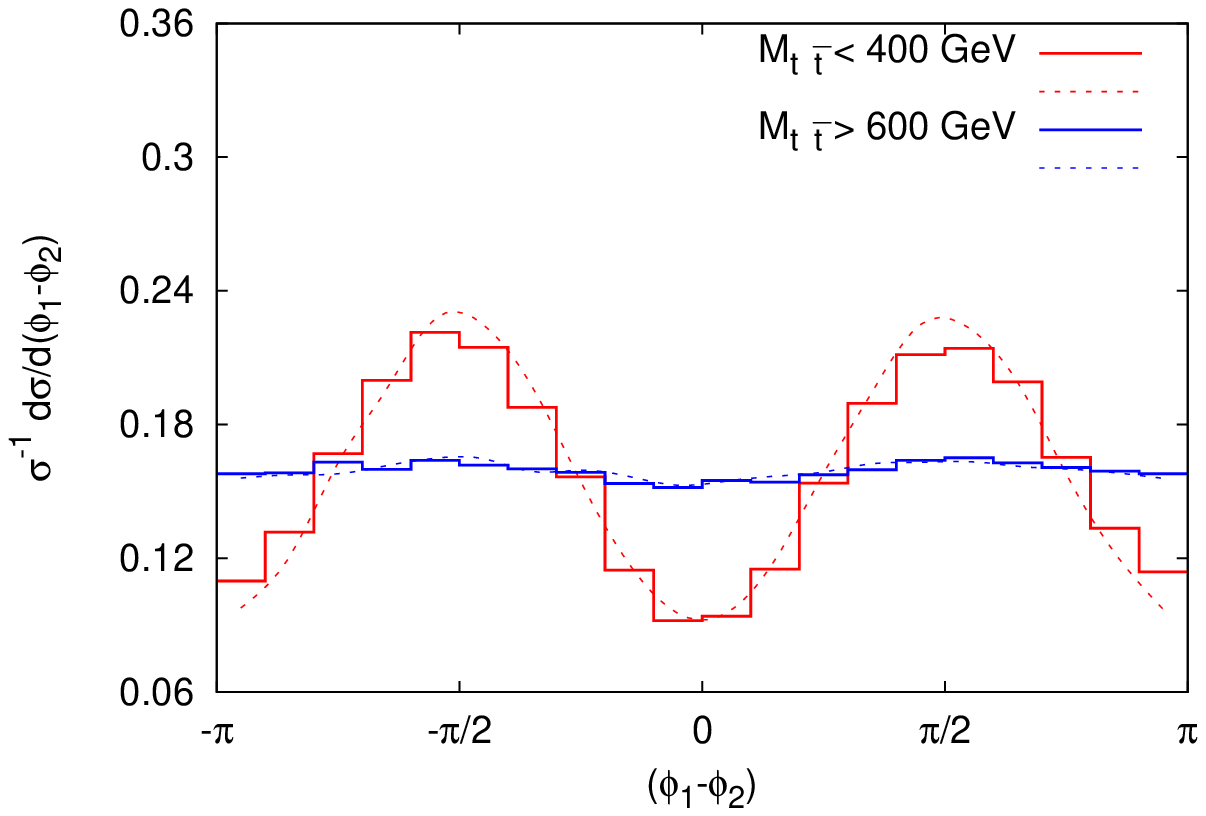,width=8.5cm,height=7cm,angle=-0}
\epsfig{file=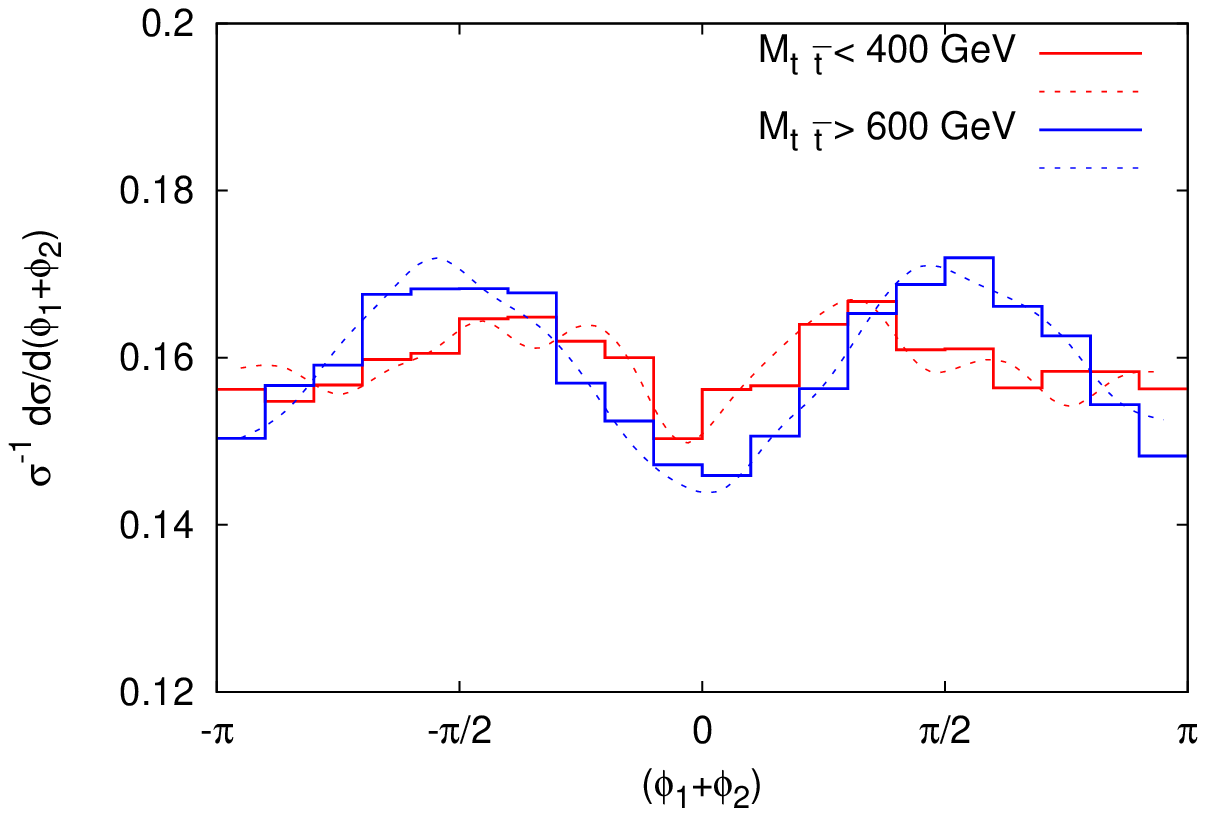,width=8.5cm,height=7cm,angle=-0}}

\caption{Normalized differential distribution of $(\phi_1-\phi_2)$ (left panel) and $(\phi_1+\phi_2)$ (right panel) for $t \bar{t}+2$~jets with different $M_{t\bar{t}}$ cuts, including all sub-processes, at the $\sqrt{s}=8 \tev$ LHC, after Cuts 1, 2 and 3 as described in the text. The distributions obtained using the exact matrix elements are shown as histograms with solid lines, while the ones obtained using the approximate matrix elements are shown as curves with dotted lines.}
\label{fig:Fig-2}
\end{center}
\end{figure}

In Figure~\ref{fig:Fig-2}, we show the normalized differential distributions in $(\phi_1-\phi_2)$ and $(\phi_1+\phi_2)$ for the $t \bar{t}+2$~jets process, where we present the sum of all sub-process contributions coming from $qq,qg$ and $gg$. Here, we have shown the effect of variation in the $M_{t\bar{t}}$ cut. While the $\phi_1-\phi_2$ distribution becomes more and more flat as we increase $M_{t\bar{t}}$, the $\phi_1+\phi_2$ distribution is observed to resemble the prediction of Eqn.~\ref{Eq:relat} for high $M_{t \bar{t}}$. It is worth mentioning that in the distributions with no cuts on $M_{t\bar{t}}$, one does retain some of the features of the two different kinematic limits. In this figure, we also compare the normalized distributions as obtained from the exact matrix elements and the approximate ones (the later being described by Eqns.~\ref{eq-Msq}-\ref{eq-Fgg}). As can be observed, the distributions very closely agree with one another, thus confirming the validity of our approximation in correctly predicting the shapes of the azimuthal angle correlations. 
\begin{table}[htb!]
\begin{center}
\begin{tabular}{|l|c c c|}
\hline
$p_T^j$ cut (GeV) &  &$\sigma_{Approx.}/\sigma_{Exact}$& \\

  &$qqt\bar{t}$ &$qgt\bar{t}$ &$ggt\bar{t}$  \\
\hline 
$20 \leq p_T^{j} \leq 60$  &2.03  &2.02 &2.06  \\
$10 \leq p_T^{j} \leq 20$  &1.08  &1.21 &1.39   \\
\hline 
\end{tabular}
\end{center}
\caption{\label{tt-comp} The ratio of $t \bar{t}+2$~jets cross-section calculated using the approximate and exact matrix elements ($\sigma_{Approx.}/\sigma_{Exact}$) for the various sub-processes, with different choices of the slicing cut for $p_T^j$, at 8 TeV LHC. All other cuts described in Cuts 1, 2 and 3 are kept fixed.}
\end{table}

We also show in Table~\ref{tt-comp} the ratio of the total cross-section obtained using the approximate and exact matrix elements. As we can see from this table, for the values of the kinematic cuts used so far in our analysis, although the on-shell gluon approximation predicts the normalized distributions rather accurately, it overestimates the total cross-section by about a factor of 2. Since our approximation approaches the exact matrix elements only in the limit when the t-channel gluons are collinear to the initial partons, we expect better agreement in the total cross-section when the $p_T^j$'s are reduced further. This is what we observe in Table~\ref{tt-comp}, where we see that by decreasing the $p_T$'s of the tagging jets, the total cross-sections also approach the exact ones. 

\begin{figure}[h]
\begin{center}
\centerline{\epsfig{file=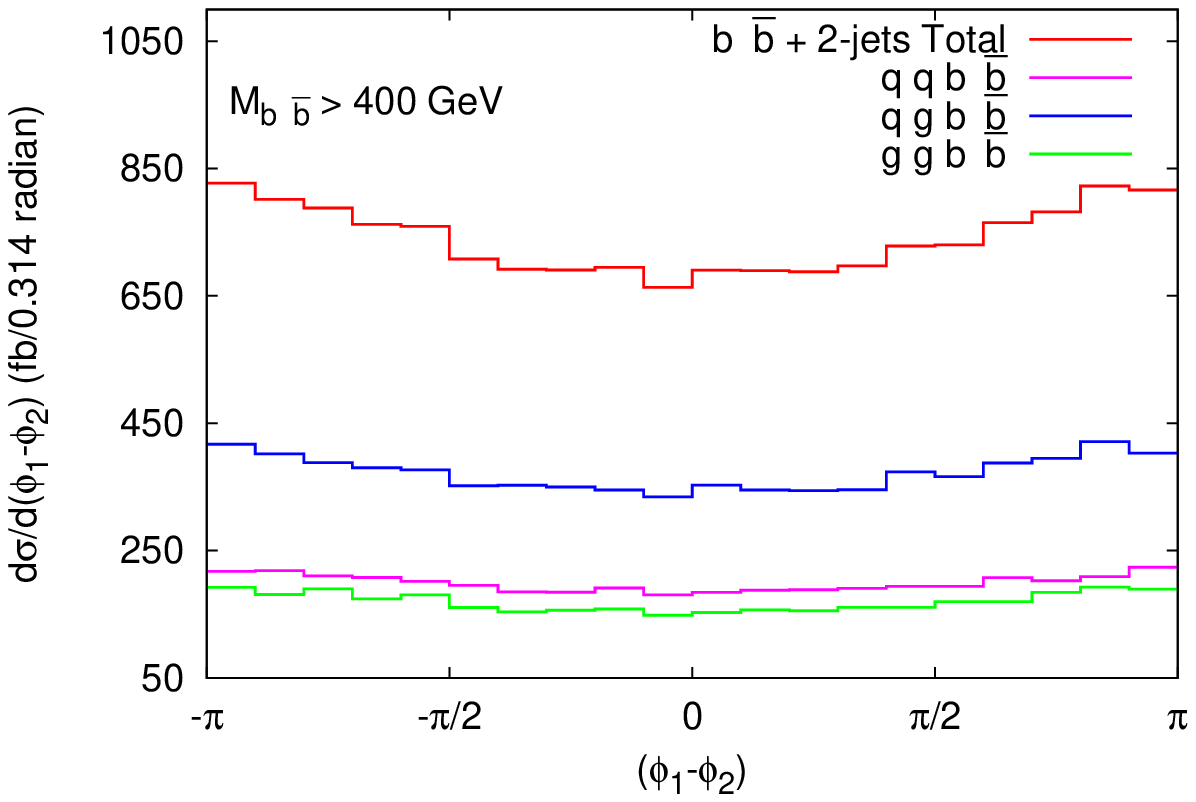,width=8.5cm,height=7cm,angle=-0}
\epsfig{file=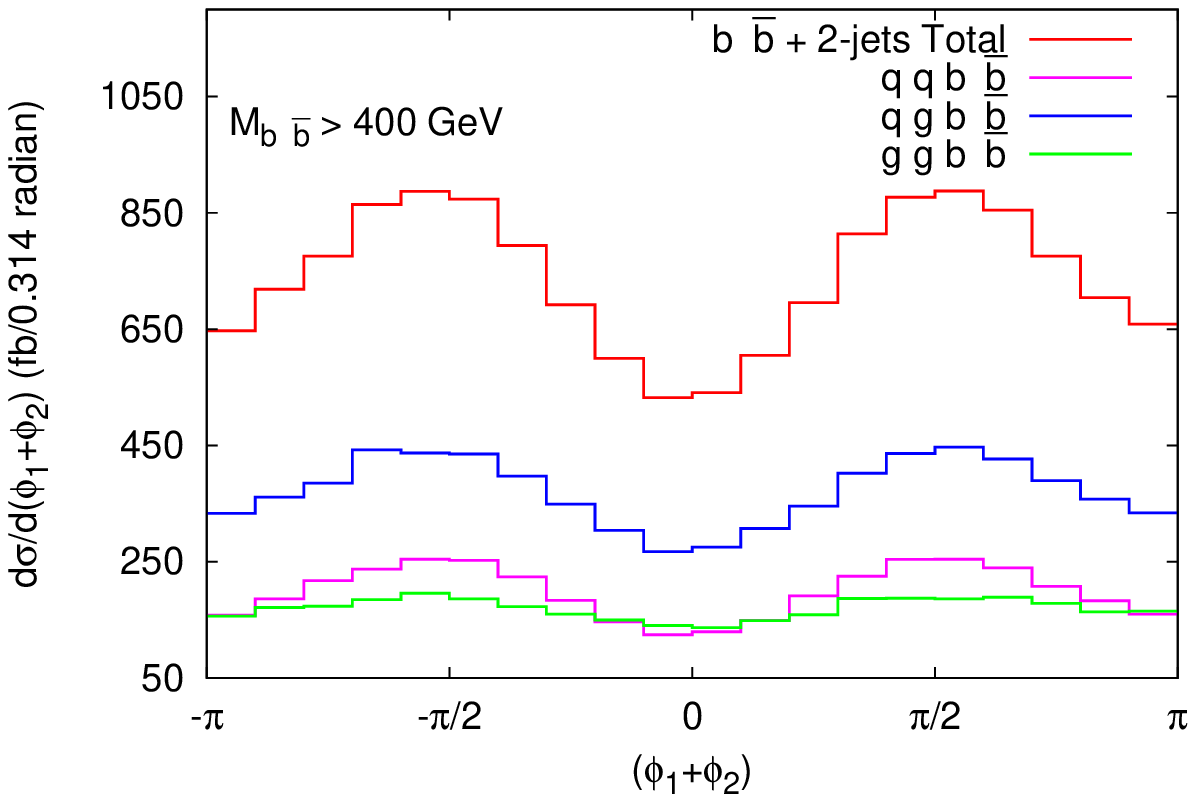,width=8.5cm,height=7cm,angle=-0}}

\caption{Differential distribution of $(\phi_1-\phi_2)$ (left panel) and $(\phi_1+\phi_2)$ (right panel) for $b \bar{b}+2$~jets in the three sub-processes initiated by $qq,qg$ and $gg$ and the sum of all sub-processes at the $\sqrt{s}=8 \tev$ LHC after Cuts 1, 2 and 3. The invariant mass of the $b \bar{b}$ pair has been demanded to be $M_{b\bar{b}} > 400 \gev$.}
\label{fig:Fig-3}
\end{center}
\end{figure}


We now turn to the $b \bar{b}+2$~jets process. Since the $b$-quark mass is very small, we can easily obtain the characteristic distributions predicted for the relativistic limit by demanding a moderate cut on $M_{b\bar{b}}$. In Figure~\ref{fig:Fig-3}, we show the differential distributions for both the azimuthal angle difference of the two tagging jets, $(\phi_1-\phi_2)$, and the sum of the azimuthal angles $(\phi_1+\phi_2)$ in $b \bar{b}+2$~jets events, for the sub-processes initiated by $qq, qg$ and $gg$. We also show the sum of all sub-process contributions separately. In this figure, we have imposed an invariant mass cut of $M_{b\bar{b}} > 400 \gev$. In contrast to $t \bar{t}+2$~jets with $M_{t \bar{t}}<400 \gev$, the $(\phi_1-\phi_2)$ distributions for $b \bar{b}+2$~jets with the chosen value of $M_{b \bar{b}}$ cut are found to be rather flat, as expected. As far as the $(\phi_1+\phi_2)$ distribution is concerned, as explained in sub-section 3.1, we indeed see a distribution as predicted by Eqn.~\ref{Eq:relat} for the $\beta \rightarrow 1$ limit. In particular, we see the expected suppression at $(\phi_1+\phi_2)=0,\pm \pi $, and enhancement at $(\phi_1+\phi_2)=\pm \pi/2 $. In order to estimate the feasibility of studying the angular distributions in $b \bar{b}+2$~jets processes at the $8 \tev$ LHC, we present in Table~\ref{Mbb} the cross-sections after Cuts 1, 2 and 3, and various values of the $M_{b \bar{b}}$ cut. The total cross-sections for this process for different values of the invariant mass cut are seen to be considerably high, such that sufficient number of events for studying the angular correlations are expected at the $8 \tev$ LHC with $\sim 20 \fb^{-1}$ luminosity, even after $b$-tagging efficiencies are taken into account.
\begin{table}[htb]
\begin{center}
\begin{tabular}{|l|c c c c|}
\hline
$M_{b \bar{b}}$ cut & $\sigma_{qqb\bar{b}}$ (fb) &$\sigma_{qgb\bar{b}}$ (fb) &$\sigma_{ggb\bar{b}}$ (fb) & $\sigma_{b\bar{b}+2~jets}$ (fb) \\
\hline 
No $M_{b \bar{b}}$ cut     &3748.27 &7374.65 &3593.10 &14716.02 \\
$M_{b \bar{b}} > 400$ GeV  &1248.36  &2333.83 &1064.38 &4646.57  \\
$M_{b \bar{b}} > 600$ GeV  &349.36  &613.72 &262.32 &1225.40  \\
\hline 
\end{tabular}
\end{center}
\caption{\label{Mbb} Cross-sections of various sub-processes contributing to $b \bar{b}+2$~jets at 8 TeV LHC after Cuts 1, 2 and 3 and different values of the $M_{b \bar{b}}$ cut. We also show the total $b \bar{b}+2$~jets cross-section after these cuts separately.}
\end{table}

\begin{figure}[h]
\begin{center}
\centerline{\epsfig{file=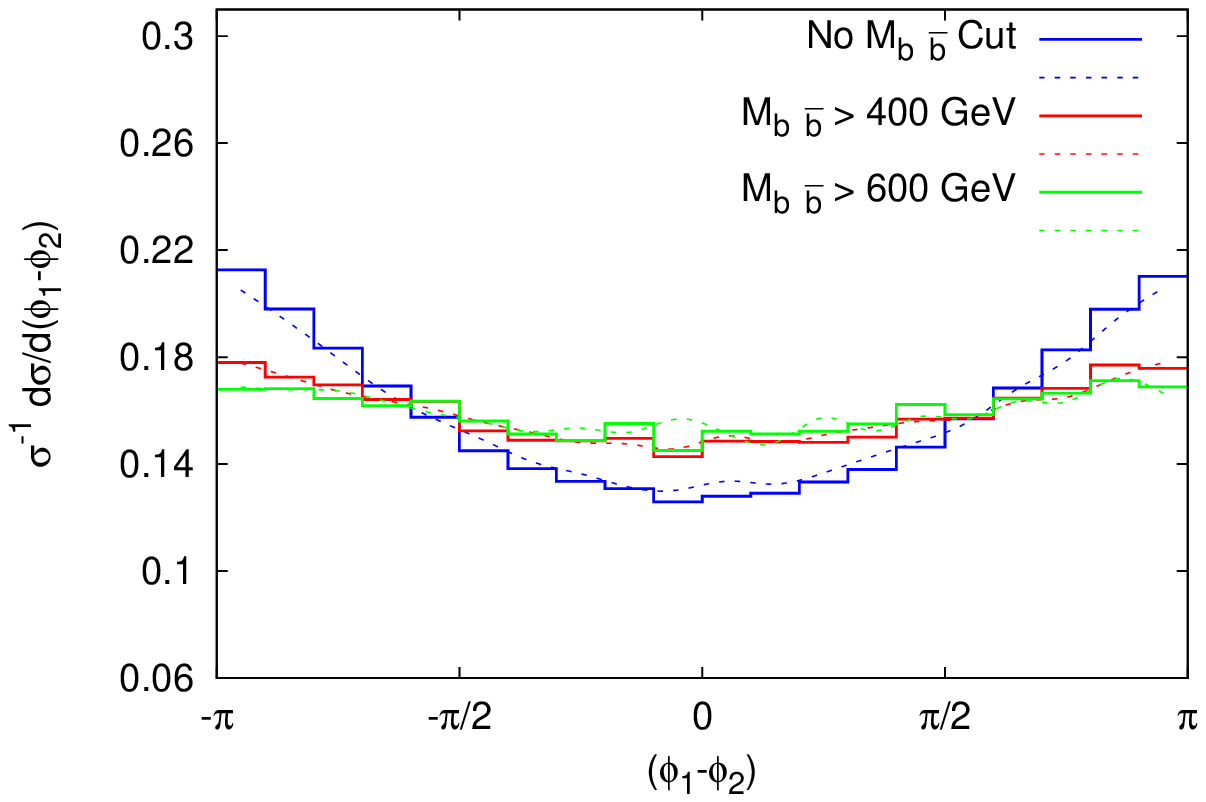,width=8.5cm,height=7cm,angle=-0}
\epsfig{file=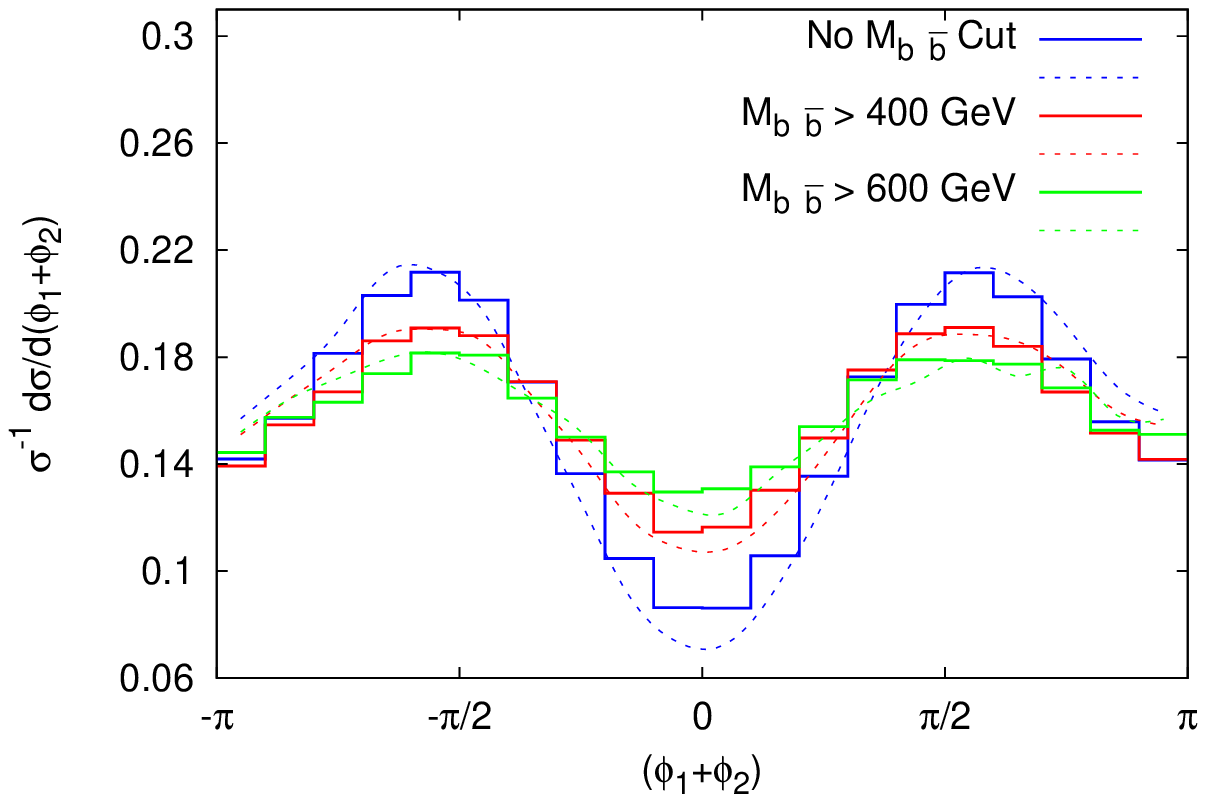,width=8.5cm,height=7cm,angle=-0}}

\caption{Normalized differential distribution of $(\phi_1-\phi_2)$ (left panel) and $(\phi_1+\phi_2)$ (right panel) for $b \bar{b}+2$~jets with different $M_{b\bar{b}}$ cuts, including all sub-processes, at the $\sqrt{s}=8 \tev$ LHC, after Cuts 1, 2 and 3. The distributions obtained using the exact matrix elements are shown as histograms with solid lines, while the ones obtained using the approximate matrix elements are shown as curves with dotted lines.}
\label{fig:Fig-4}
\end{center}
\end{figure}

\begin{table}[htb!]
\begin{center}
\begin{tabular}{|l|c c c|}
\hline
$p_T^j$ cut (GeV) &  &$\sigma_{Approx.}/\sigma_{Exact}$& \\

  &$qqb\bar{b}$ &$qgb\bar{b}$ &$ggb\bar{b}$  \\
\hline 
$20 \leq p_T^{j} \leq 60$  &2.34  &2.31 &2.27  \\
$10 \leq p_T^{j} \leq 20$  &2.07  &2.06 &2.07   \\
\hline 
\end{tabular}
\end{center}
\caption{\label{bb-comp} The ratio of $b \bar{b}+2$~jets cross-section calculated using the approximate and exact matrix elements ($\sigma_{Approx.}/\sigma_{Exact}$) for the various sub-processes, with different choices of the slicing cut for $p_T^j$, at 8 TeV LHC. All other cuts described in Cuts 1, 2 and 3 are kept fixed.}
\end{table}

In Figure~\ref{fig:Fig-4}, we show the normalized differential distributions for the $b \bar{b}+2$~jets process, where we present the sum of all sub-process contributions coming from $qq,qg$ and $gg$. Here, we have shown the effect of variation in the $M_{b\bar{b}}$ cut. We also compare the normalized distributions obtained by using the approximate and the exact matrix elements, and they are found to agree to a good accuracy, as in the $t \bar{t}+2$~jets case. We show in Table~\ref{bb-comp} the ratio of the total cross-section obtained by using the exact and approximate matrix elements for two different choices of the slicing $p_T$ cuts on the tagging jets. Here also, the approximate matrix element is found to overestimate the total cross-section, and the difference between the two is reduced once the tagging jet $p_T$'s are demanded to be very small, thereby approaching the collinear limit for the t-channel gluons.

\section{Summary} 
In this paper, we have studied the azimuthal angle correlations of jets produced in association with a top or bottom quark pair at the LHC. We have presented the helicity amplitudes for the process $Q \overline{Q}+2$~jets using only the gluon fusion diagrams and employing the on-shell gluon approximation to the matrix element. This gives us a general expression for the azimuthal angle distributions for processes initiated by $qq, qg$ and $gg$ initial states. Subsequently, we have explored this matrix element in two different kinematic limits. The first one is the threshold production of the $Q \overline{Q}$ pair, in which the invariant mass $M_{Q \overline{Q}}$ is very near to $2 m_Q$. In this limit we find a strong angular correlation in the distribution of the difference of azimuthal angles of the tagging jets, $(\phi_1-\phi_2)$. The other kinematic limit considered is the relativistic limit, in which $M_{Q \overline{Q}}$ is very much higher than $2 m_Q$. In this limit, while the distribution in $(\phi_1-\phi_2)$ is expected to be rather flat, there is a strong correlation in the sum of the azimuthal angles, $(\phi_1+\phi_2)$. We performed an exact numerical evaluation of these angular distributions in $t \bar{t}+2$~jets and $b \bar{b}+2$~jets processes at the LHC, including all the diagrams at the tree level, and verified our analytical approximations using a suitably devised set of kinematic selection cuts. The VBF selection cuts are found to efficiently select the GF contribution to these processes. A comparison between the normalized differential distributions obtained by using the exact and the approximate matrix elements also shows a very good agreement between the two. The total cross-sections are overestimated by the approximate matrix elements by about a factor of 2, while they approach the exact cross-sections as the $p_T$'s of the tagging jets are demanded to be smaller, thereby achieving the collinear limit for the t-channel intermediate gluons, in which our approximation approaches the exact matrix elements. 

We expect these azimuthal correlations calculated at the leading order to survive even after including higher order corrections, as has been previously demonstrated in the case of Higgs production with multiple hard jets~\cite{HiggsCP,higher-order}. One particular future direction that can be explored is to extend our parton level results by using a matrix elements matched with parton shower method. 

We would like to stress that the study of such angular correlations among jets produced with a top or bottom quark pair is not only interesting in itself, it has useful implications in the study of VBF processes for determining the spin and CP properties of new particles which can be discovered at the LHC. In particular, the experimental technique to measure such correlations between tagging jets can be established first by using these SM processes which have sufficiently high cross-sections. We expect that these studies can be performed with the already accumulated data at the 8 TeV run of the LHC.
\section*{Acknowledgements} 
SM is grateful to Junichi Kanzaki for many a help regarding the use of the {\tt BASES} integration package. SM would also like to acknowledge the warm hospitality of the Theory Center at KEK, Japan, and the RECAPP centre at Harish-Chandra Research Institute, India, where major portions of this work were carried out. This work is partially supported by World Premier International Research Center Initiative (WPI Initiative), MEXT, Japan, and by Grant-in-Aid for scientific research (No.-20340064) from JSPS.
\vspace{0.5cm}


\end{document}